\documentstyle[bo99,epsfig]{article}

\title{New results from the HRX BL Lac sample}
\author{Volker Beckmann$^{1,2}$ and Anna Wolter $^2$}
\affil{1) Hamburger Sternwarte, Gojenbergsweg 112, 21029 Hamburg, Germany\\
2) Osservatorio Astronomico di Brera, Via Brera 28, 20121 Milano, Italy}

\begin{document}

\maketitle

\begin{abstract}
We present results for the Hamburg BL Lac sample, based on data
provided by the RASS-BSC. 
By fitting a single power law to the X-ray data we find, in a number of 
objects, an additional absorbing component to the galactic value of $N_{H}$, 
which might be attributed to intrinsic absorption.  A more probable cause
seems however to be a curvature in the X-ray spectra in the sense that 
they are more curved for steeper slopes.  
The known relation between the X-ray spectral
slope and the ratio between optical and X-ray flux ($\alpha_{OX}$)
also applies to this BL Lac sample, even though less significant than
in previous works. We also find a dependence of X-ray luminosity on
$\alpha_{OX}$.
\end{abstract}

\section{Introduction}
BL Lac objects are rare AGN, which are thought to be oriented towards
us with their jet, thus showing high polarisation, strong variability
at all wavelengths and non-thermal featureless spectra.  Still there
are several open questions about the nature of the BL Lac objects,
e.g. whether there are differences between the BL Lac objects which are
selected due to their strong radio-emission (RBL) and those selected
because of their X-ray brightness (XBL). The difference between those
two classes might be mainly caused by different peak frequencies of
the two components in the BL Lac spectrum: XBL have usually a higher
peak frequency of the synchrotron branch than RBL and thus we have
high-frequency (HBL) and low frequency (LBL) peaked BL Lac objects
(see e.g. Fossati et al. 1998). Between those two classes one can find
intermediate objects (IBL) whose properties show them to be the link
between HBL and LBL. We study the X-ray spectra of IBL and HBL and fit 
them with a single-power law with absorption by neutral hydrogen to 
investigate any possible spectral differences between these two
samples.

\section{The Hamburg X-ray bright BL Lac sample}
This work is based on the bright sources from ROSAT All Sky Survey
(Voges et al. 1996), which have an X-ray flux $f_X > 11\cdot 10^ {-13}
{\rm erg \; cm^{-2} \; sec^{-1}}$ in the hard PSPC energy band ($0.5 -
2.0$ keV). A first complete sample of 39 BL Lac objects (Bade et
al. 1998) in an area of $2800 \; {\rm deg}^{2}$ is based on the
Hamburg/RASS X-ray bright sample (HRX, Cordis et al. in preparation).  
A larger sample of candidates has been derived by cross-correlating
the X-ray sources from RASS in a larger area ($4500 \; {\rm deg}^{2}$)
with radio catalogues (NVSS, FIRST: radio flux limit
$\simeq 2.5 \; {\rm mJy}$).
These 262 sources, if not already classified, have been included in
follow-up spectroscopy with the 3.5m telescope on Calar Alto. Up to
now this statistically complete sample is \mbox{95 \%} classified and
contains 72 BL Lac objects, both of the HBL and the IBL type.  
For 61 of them we have
already determined the redshift ($z < 0.9$; \mbox{Beckmann 1999}). We
will use in the following this sample of 72 BL Lacs.

\section{X--ray spectra}
Since the only available spectral measures are the two hardness ratios 
given in the RASS-BSC, we can only assume a simple spectral shape and
apply the method of Schartel et al. (1996) to determine its parameters.
We therefore assume that the spectrum is in the form of
a single-power law absorbed by two different contributions: the galactic 
absorption and an ``intrinsic'' absorption that is given by the 
difference between the best fit absorption and the Galactic one
($N_{H,intrinsic} = N_{H,free-fitted} - N_{H,galactic}$). 
The galactic values were taken from Dickey and
Lockman (1990).  We find a strong correlation of the
``intrinsic'' absorption with the spectral slope (Fig. 1) on a $> 99 \%$
confidence level also when taking into account the large errors in
$N_{H,intrinsic}$.
\begin{figure}
\centerline{\psfig{file=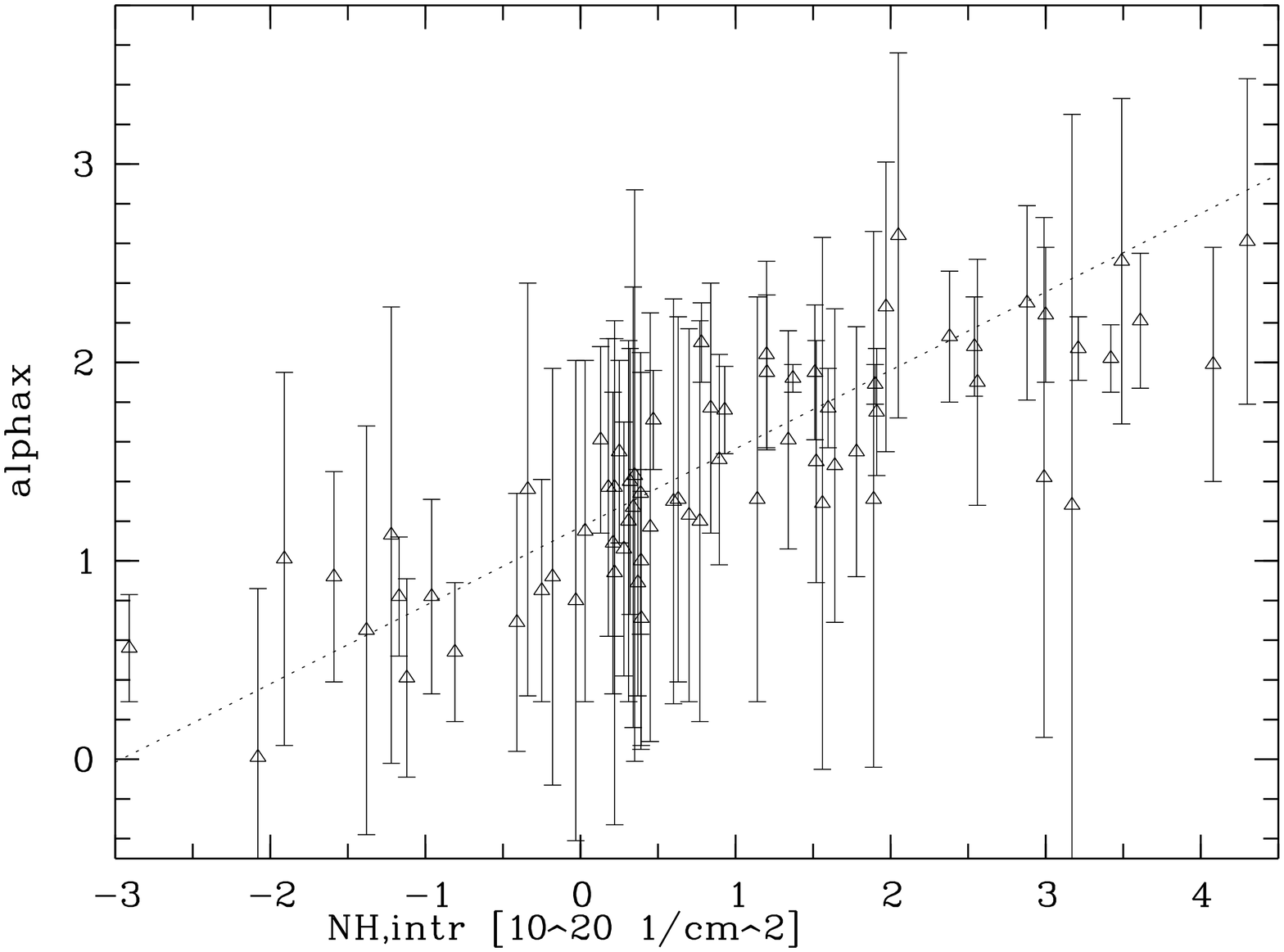, width=6.5cm}}
\caption[1]{The ``intrinsic'' absorption ($N_{H,intrinsic} = N_{H,free-fitted} 
- N_{H,galactic}$) versus X-ray spectral slope. The negative values
are caused by large errors on the free-fitted $N_{H}$ so that they are
consistent with $N_{H,intrinsic} = 0$. The linear regression takes the
errors in $N_{H}$ and $\alpha_{X}$ into account.}
\end{figure}
The negative values of $N_{H,intrinsic}$ seem to be caused by large
errors; if we just take into account objects with an error of
$N_{H,intrinsic} < 0.3$, we do not have any ``negative absorption'' at
all.  Since we do not see any correlation to other observable
parameters (such as flux or luminosity), we make the hypothesis that
this is an effect of curvature in the X-ray spectra (see e.g. Urry et
al. 1996).  Thus the flat X-ray spectra are well described by a single
power law with galactic low energy absorption ($N_{H,intrinsic} \simeq
0$), while for steep spectra the curvature is significant and an
additional ``absorption'' is needed to fit the X-ray data to a
single-power law. In this model, a large value of $N_{H,intrinsic}$
would be explained by a convex spectrum. Nevertheless, true intrinsic
absorption can not be ruled out in principle.

\section{$\alpha_{OX}$ - $\alpha_{X}$ relation for HRX-BL Lac}
A relation, which is found for BL Lac objects, is the dependence of
the spectral slope $\alpha_{X}$ on $\alpha_{\rm OX}$\footnote{we
define $\alpha_{\rm OX}$ as the power law index between $1$ keV and
$4400$ {\AA} with $f_{\nu} \propto \nu^{-\alpha_{\rm OX}}$, that
describes the X-ray dominance over the optical brightness}.  This
effect, which shows objects with a higher X-ray dominance having
flatter X-ray spectra, was detected in several X-ray bright samples of
BL Lac objects (e.g. Wolter et al. 1998, Padovani \& Giommi 1995,
Comastri et al. 1995).\\
Based on the RASS data we determined this relation for the HRX-BL
sample (Fig.2). The effect of spectral flattening in the X-ray region
with decreasing $\alpha_{\rm OX}$ is detectable, even though less
significant than in previous works. An analysis, which takes into
account the errors in $\alpha_{\rm OX}$ and $\alpha_{\rm X}$ gives a
significance for correlation of both values on a $>93 \%$ level.
\begin{figure}
\centerline{\psfig{file=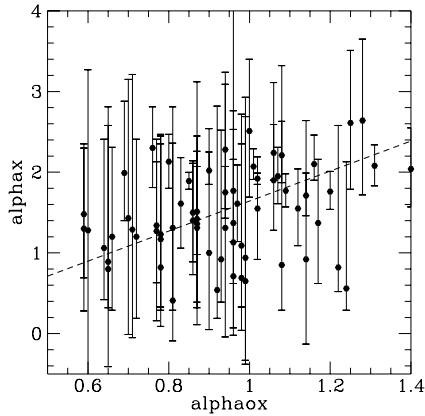, width=6.0cm}}
\caption[]{The relation between X-ray dominance ($\alpha_{OX}$) and X-ray
  spectral slope ($\alpha_{X}$) for the whole BL Lac sample. X-ray
  dominant objects are on the left.}
\end{figure}

\section{Dependence of X-ray luminosity on $\alpha_{OX}$}
Another physical parameter which seems to be correlated to the X-ray
dominance is the X-ray luminosity of the BL Lac objects.  Because we
know the redshift and thus the luminosity for more than $80 \%$ of our
objects, it is possible to study this relation.  Figure 3 shows the
dependency of $\log L_{X}$ on $\alpha_{OX}$. There seem to be no
objects, which are X-ray dominant (low $\alpha_{OX}$) and have a low
X-ray luminosity. This result is solid, because we do not miss
any redshift for objects with $\alpha_{OX} < 0.9$ (Beckmann 1999).
This effect could be explained in the view of the unified schemes
(e.g. Padovani \& Giommi 1995). The HBL are more X-ray dominated (low
$\alpha_{OX}$) and the X-ray luminosity is therefore high because the
X-ray band is in the vicinity of the peak frequency.  On the other
hand, LBL (high $\alpha_{OX}$) have their maximum of the synchrotron
emission in the optical region and are X-ray faint.  
\begin{figure}
\centerline{\psfig{file=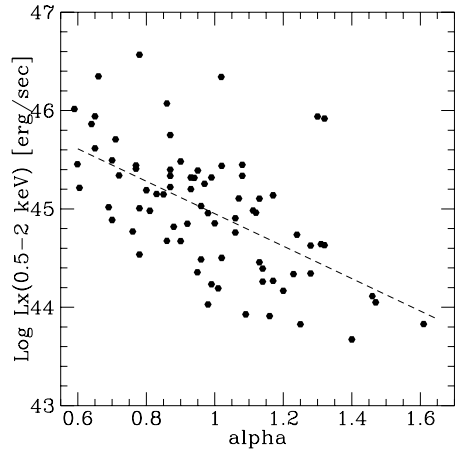, width=6.5cm}} 
\caption[]{The
relation between X-ray dominance ($\alpha_{OX}$) and X-ray luminosity
in the hard ($0.5 - 2 {\rm keV}$) ROSAT-PSPC band. There are no X-ray
faint BL Lacs with low $\alpha_{OX}$.}  
\end{figure}

\end{document}